\documentclass[twocolumn,showpacs,preprintnumbers,amsmath,amssymb,pra]{revtex4}
\usepackage{epsfig}
\usepackage{bm}
\usepackage{epstopdf}

\begin{document}

\title{Mapping the Two-Component Atomic Fermi Gas to the Nuclear Shell-Model}

\author{C. {\"O}zen}
\affiliation{Center for Theoretical Physics, Sloane Physics Laboratory,Yale University, New Haven, CT 06520, USA}
\affiliation{Faculty of Engineering and Natural Sciences, Kadir Has University, Istanbul 34083, Turkey}
\author{N.~T. Zinner}
\affiliation{ Department of Physics, Harvard University, MA 02138, USA}
\affiliation{ Department of Physics and Astronomy, Aarhus University, DK-8000 Aarhus C, Denmark}

\date{\today}

\begin{abstract}
The physics of a two-component cold fermi gas is now frequently addressed in laboratories. 
Usually this is done for large samples of tens to hundreds of thousands of particles. 
However, it is now possible to produce few-body systems (1-100 particles) in very tight traps where the 
shell structure of the external potential becomes important. 
A system of two-species fermionic cold atoms with an attractive zero-range interaction is analogous to 
a simple model of nucleus in which neutrons and protons interact only through a residual pairing interaction.
In this article, we discuss how the problem of a two-component atomic fermi gas in a tight external trap can be mapped 
to the nuclear shell model so that readily available many-body techniques in nuclear physics, such as the
Shell Model Monte Carlo (SMMC) method, can be directly applied to the study of these systems.
We demonstrate an application of the SMMC method by estimating the pairing correlations in a small two-component Fermi system with 
moderate-to-strong short-range two-body interactions in a three-dimensional harmonic external trapping potential. 

\end{abstract}

\maketitle

\section{Introduction}
The physics of ultracold gases has seen a rapid development over the past decade \cite{bloch2008,giorgini2008,ketterle2008}. 
An interesting  goal in the boundary of few- and many-body systems is the implementation of optical microtraps that 
can hold a small number of particles. This was recently achieved by the Jochim group in Heidelberg \cite{selim1,selim2}. 
These experiments were performed in a regime where the trapping shell structure became prominent. 
In theoretical investigations of these systems, large quantum fluctuations would invalidate the use of mean-field 
approaches such as the BCS method due to the small number of particles involved; hence, many-body approaches beyond the mean-field are needed. 
Also, the problem of a small number of fermions interacting with each other 
in the presence of external fields---which provide a level structure---
is very similar to the nuclear pairing problem which was initially described in the seminal work of Bohr, Mottelson, and Pines \cite{bohr1958}.
There is thus a strong incentive to transfer methods from nuclear physics into cold atomic gases~\cite{carlson2012,zinner2013}.

In this article we outline the mapping of a two-component atomic fermi gas, confined by a tight external trap and interacting through a zero-range 
interaction, onto the nuclear shell-model in detail. 
For the benefit of both nuclear and atomic physics communities, 
the explicit evaluation of the matrix elements of the zero-range interaction in  
the nuclear shell model is given in considerable detail in the appendix. 
We then investigate 
the pairing correlations of small systems (less than 20 particles) using the Shell Model Monte Carlo (SMMC) 
technique that has been succesful
in nuclear physics~\cite{lang1993,alhassid1994,koonin1997}.
Also we briefly comment on some of the similarities and 
differences in studies of the pairing phenomena in the fields of atomic and nuclear physics. 
Our discussion of the pairing correlations through a two-body BCS-like pairing matrix,
to the best of our knowledge, has not been considered in the context of small ultracold Fermi systems in traps before.

A number of different approaches have been used in recent years to address the energetics, structure, and other
properties of small Fermi systems~\cite{chang2007,blume2007, forbes2012, carlson2012,blume2012}.  
Although we will briefly comment on and relate to these developments as we proceed, the purpose of this study 
is not to make a detailed quantitative comparison of various methods in use. Instead, we describe the technical issue of 
mapping the atomic gas problem in order to apply the SMMC method, a traditional nuclear physics tool, in studies of small ultracold Fermi 
systems.  On a side note, the mapping  could also provide a natural connection between
atomic and nuclear physics given the prospect that ultracold atomic systems with spin-orbital 
momentum coupling---a central tenet in nuclear shell model---may soon be realized.

\section{Mapping the Fermi Gas to the Nuclear Shell Model}
The two-component ultracold fermionic atomic gas consists of neutral atoms, usually alkali species, 
that occupy two different internal states. 
The actual internal states are hyperfine states of different projection that 
can be split by a magnetic 
field \cite{ketterle2008}. The energy scale of the hyperfine splitting is by far larger than any other energy scale in the problem so that no internal process in the gas can transfer atoms between the hyperfine levels; thus, one may think of these levels as frozen degrees of freedom. Also; since these systems are usually dilute, the range of the atom-atom interactions is very short compared to the typical interparticle distance. Therefore, the simple zero-range potential is a popular and highly successful model.

The three-dimensional $N$-body Fermi system in an isotropic harmonic trap with a zero-range interaction of strength $V_0$ can be described by the Hamiltonian
\begin{equation} \label{ham}
H=\sum_i \frac{\vec{p}_{i}^{2}}{2m}+\sum_i \frac{1}{2}m\omega^2  \vec{r}_{i}^{2}+\sum_{[ij]}V_0
\delta(\vec{r}_i-\vec{r}_j),
\end{equation}
where $i,j$ denote the particles, $[ij]$ denotes the sum over all pairs of particles, $m$ is the mass of the particles, and $\omega$ is the 
external trapping angular frequency. The oscillator length which we will use later is given by $b=\sqrt{\hbar/m\omega}$.
We note that the isotropic three-dimensional oscillator 
potential has shell closures at $N=2$, 8, and 20 (the $s$, $s+p$, and $s+p+sd$ shell configurations in 
typical nuclear physics language). These will be prominent features in our examples later.
In the following, we use the notation where the matrix elements of a general two-body interaction $V_{int}$ are given by
\begin{equation}
\langle \psi_a(\vec{r}_1)\psi_b(\vec{r}_2) | V_{int} | \psi_c( \vec{r}_1 )\psi_d( \vec{r}_2 )\rangle,
\end{equation}
in which the two-body wave functions $\psi_a(\vec{r}_1)\psi_b(\vec{r}_2)$ and $\psi_c(\vec{r}_1)\psi_d(\vec{r}_2)$ must be antisymmetric under the  exchange of coordinates.

In a tight harmonic trap (small trapping length, $b=\sqrt{\tfrac{\hbar}{m\omega}}$), 
the quantum numbers of single-particle levels are 
given by $n,l,m_l$. The two internal hyperfine states can now be mapped onto 
the spin of a single species of nucleon $m_s=\pm 1/2$. Thus, any single-particle 
state is uniquely described by $a=(n_a l_a m_{l_a} m_{s_a})$. Any two-body 
state constructed from these states will have an external and an internal part 
that combine to determine the overall symmetry. We use a zero-range 
interaction and the spatial part must thus be non-zero at the origin to give a 
contribution. This is only possible with a relative wave function that 
is symmetric under particle exchange. Since the particles are fermions,
the internal (hyperfine, or pseudospin) state must 
be antisymmetric, or in the spin $1/2$ language, a spin-singlet state. 
This completes our mapping of the two-component fermi system in a trap onto the 
nuclear shell model with one species of nucleon (proton or neutron). The nuclear 
mean-field is replaced by the harmonic oscillator and the internal spin 
states of the nucleon now correspond to the hyperfine states for the atoms.

In general, the three-dimensional zero-range interaction is
ill-defined unless properly regularized. As it has been shown by Busch {\it et al.}
\cite{busch98}, the case of two fermions with different internal states in a 
harmonic potential interacting via a zero-range interaction cannot only be 
properly regularized, but in fact has a tractable solution. This 
solution has been subsequently studied and confirmed by atomic physics 
experiments \cite{stoferle06}. 
In relation to shell-model applications,
the issue is always that a finite model space is used. 
However, having access to the exact solution in the full space
of Busch {\it et al.} is an excellent starting point for doing 
many-body problems in both nuclear and atomic physics \cite{haxton02}.
In the case of the SMMC method that we 
are concerned with here, the question of regularization was 
discussed in detail in Ref.~\cite{zinner2008}. 

Below we 
will be using strengths $g=-V_0/(\hbar\omega b^3)=10$ 
and $g=20$. The strength can also be given in terms 
of the two-body scattering length, $a$. For $g=10$
we have $a/b=-1.0$ and for $g=20$ we have $a/b=11$.
Comparing to the standard usage in BCS-BEC 
crossover studies \cite{bloch2008,giorgini2008}
the first value $a/b=-1.0$ is on the (deep) BCS side, while
the $a/b=11$ value is on the BEC side but close to the 
resonance ($a\to\infty$) and thus close to the 
unitarity limit.

\section{Shell Model Monte Carlo Method}

Quantum Monte Carlo methods have been extensively used in the study of strongly interacting many-body problems (see f.x. Ref.~\cite{carlson2012} 
and references therein). An example is the Auxilary-field Monte Carlo (AFMC) approach of Zhang and collaborators~\cite{zhang1995,zhang2003}. 
The AFMC method have been used to calculate zero- and finite-temperature properties of the unitary Fermi gas on a lattice~\cite{lee2006,bulgac2006}.
As an alternative to the lattice representation, the AFMC is also formulated within the configuration-interaction nuclear shell-model. 
This approach is known as the the Shell-Model Monte Carlo (SMMC) method and has been widely employed in nuclear physics~\cite{lang1993,alhassid1994,koonin1997} and more recently in the study of trapped cold atoms~\cite{zinner2008,chris2012,mukherjee2013}.
The SMMC approach is based on a linearization
of the two-body part of the Hamiltonian using the Hubbard-Stratonovich transformation~\cite{hubbard1959}. Here we adopt a formulation of this
transformation starting from a general Hamiltonian, which can be written in a manifestly time-reversal invariant form:
\begin{equation}
H=\sum_\alpha \left(\epsilon_\alpha \mathcal{O}_\alpha +\epsilon_{\alpha}^{*} \bar{\mathcal{O}}_{\alpha}\right)+\frac{1}{2}\sum_\alpha V_\alpha
\left\{ \mathcal{O}_\alpha,\bar{\mathcal{O}}_{\alpha} \right\},
\label{Eq:GenHam}
\end{equation}
where $O_\alpha$ are one-body operators in a convenient basis and the $V_\alpha$ are real numbers. The bars denote time-reversed operators. 
The SMMC approach relies on the Hubbard-Stratonovich (HS) transformation to linearize the many-body evolution operator $e^{-\beta H}$, where $\beta^{-1}$ may be interpreted as the temperature in the (grand)canonical ensemble. We first divide $\beta$ into $N_t$ time slices so that we can express individual terms at different time slices in $e^{-\beta H}=\left[ e^{-\Delta\beta H} \right]^{N_t}$ as 
\begin{widetext}
\begin{eqnarray}
e^{-\Delta\beta H} \approx e^{-\Delta\beta\sum_\alpha \left(\epsilon_\alpha \mathcal{O}_\alpha +\epsilon_{\alpha}^{*}    
  \bar{\mathcal{O}}_{\alpha}\right) } \prod_{\alpha} e^{-\Delta\beta\frac{V_\alpha}{4} \left[ 
  (\mathcal{O}_{\alpha}+\bar{\mathcal{O}}_{\alpha})^2  - (\mathcal{O}_{\alpha}-\bar{\mathcal{O}}_{\alpha})^2 \right] }
  + \mathcal{O}(\Delta\beta)^2 \nonumber,
\end{eqnarray}
\end{widetext}
where we used $2\left\{ \mathcal{O}_\alpha,\bar{\mathcal{O}}_{\alpha} \right\}=(\mathcal{O}_{\alpha}+\bar{\mathcal{O}}_{\alpha})^2  - (\mathcal{O}_{\alpha}-\bar{\mathcal{O}}_{\alpha})^2$. Quadratic interaction terms can be effectively linearized through the Gaussian integral identity
\newpage
\begin{widetext}
\begin{eqnarray}
   e^{-\Delta\beta\frac{V_\alpha}{4} \left[ (\mathcal{O}_{\alpha}+\bar{\mathcal{O}}_{\alpha})^2  - 
     (\mathcal{O}_{\alpha}-\bar{\mathcal{O}}_{\alpha})^2 \right] } = \frac{\Delta\beta\vert V_\alpha\vert}{4\pi} \int 
     d\sigma_\alpha^R
     d\sigma_\alpha^I e^{-\Delta\beta\frac{\vert V_\alpha \vert}{4}\left[ (\sigma_\alpha^R)^2+(\sigma_\alpha^I)^2\right]} \nonumber 
     \\
     e^{-\Delta\beta\frac{V_\alpha}{2}\left[ s_\alpha \sigma_\alpha^R  (\mathcal{O}_{\alpha}+\bar{\mathcal{O}}_{\alpha})
        +i s_\alpha \sigma_\alpha^I  (\mathcal{O}_{\alpha}-\bar{\mathcal{O}}_{\alpha}) \right]},
\end{eqnarray}
\end{widetext}
where the integration variables $\sigma_\alpha^R$ and $\sigma_\alpha^I$ are the real auxiliary fields that give the method its name.
The sign factors are $s_\alpha=\pm 1$ for $V_\alpha<0$ and $s_\alpha=\pm i$ for $V_\alpha>0$.
Introducing complex fields for each time slice $\sigma_\alpha(\tau_n)=\sigma_\alpha^R(\tau_n)+i\sigma_\alpha^I(\tau_n)$, we arrive at the Hubbard-Stratonovich representation of the many-body evolution operator 
\begin{eqnarray}
  e^{-\beta H}=\int \mathcal{D}[\sigma] G(\sigma) U_\sigma(\beta,0).
\end{eqnarray}
Above, 
\begin{equation}
   \mathcal{D}[\sigma]= \prod_{\alpha,n} \frac{d\sigma_\alpha(\tau_n) d\sigma^*_\alpha(\tau_n)}{2 i}\frac{\Delta\beta\vert V_\alpha \vert}{4\pi}
\end{equation}
is the measure of the integral. $G(\sigma)$ is a Gaussian weight
\begin{equation}
    G(\sigma)=e^{-\frac{\Delta\beta}{4}\sum_\alpha \vert V_\alpha \vert \vert \sigma_\alpha(\tau_n) \vert^2}.
\end{equation}
The Many-body propagator, $e^{-\beta H}$, is now effectively reduced to a superposition of one-body propagators 
\begin{equation}
 U_\sigma(\beta,0)=e^{-\Delta\beta h_\sigma(\tau_{N_t})} \ldots e^{-\Delta\beta h_\sigma(\tau_{1})},
\end{equation}
where the linearized Hamiltonian as a function of the time-dependent auxiliary fields is given by
\begin{widetext}
\begin{equation}
  h_\sigma(\tau)=\sum_\alpha \left(\epsilon_\alpha  + \frac{1}{2}  s_\alpha V_\alpha \sigma_\alpha(\tau) \right)  \mathcal{O}_\alpha
    +            \left( \epsilon_{\alpha}^{*} + \frac{1}{2}  s_\alpha V_\alpha \sigma_\alpha^{*}(\tau) \right) \bar{\mathcal{O}}_{\alpha}.
  \label{Eq:LinHam}
\end{equation}
\end{widetext}
In the SMMC, expectation value of an observable $\Omega$ at temperature $T=1/\beta$ is calculated by expressing both the numerator and the denominator of $\langle \Omega \rangle=\mathrm{Tr}_N[\Omega e^{-\beta H}]/\mathrm{Tr}_N e^{-\beta H}$ (where $\mathrm{Tr}_N$ denotes a canonical trace for N-particle system) in the HS representation. In order to perform a Monte Carlo integration, a positive definite weight function is defined as $W(\sigma)=G(\sigma)\vert \mathrm{Tr}_N U_\sigma(\beta,0) \vert$. Thus; one can express the thermal expectation values by 
\begin{eqnarray}
   \langle \Omega \rangle
                         = \frac{\int \mathcal{D}[\sigma] W(\sigma) \Phi(\sigma) \langle \Omega \rangle_\sigma}
                           {\int \mathcal{D}[\sigma] W(\sigma) \Phi(\sigma) }, 
\end{eqnarray}
where $\Phi(\sigma)=\mathrm{Tr}_N U_\sigma(\beta,0)/\vert \mathrm{Tr}_N U_\sigma(\beta,0) \vert$ is the ``sign'' and 
$\langle \Omega \rangle_\sigma=\mathrm{Tr}_N \vert \Omega U_\sigma(\beta,0)\vert / \mathrm{Tr}_N U_\sigma(\beta,0)$. 
The observable $\langle \Omega \rangle$ is then computed in a Monte Carlo integration by selecting an ensemble of auxiliary fields ($\sigma_1, \ldots, \sigma_N$)
sampled according to the distribution function $W(\sigma)$, i.e.,  
\begin{equation}
 \label{Eq:MonteCarlo}
 \langle \Omega \rangle \approx  \frac{\frac{1}{N} \sum_n \Phi (\sigma_n) \langle \Omega \rangle_{\sigma_n}}
                            {\frac{1}{N} \sum_n \Phi (\sigma_n)}.
\end{equation}
Success of the outlined method hinges on the sign $\Phi(\sigma)$ of the weight function $W(\sigma)$. 
Unfortunately, in the most general case, $\mathrm{Tr}_N U_\sigma(\beta,0)$ is not always positive 
hence $\Phi(\sigma)$ can be $\pm 1$. Such fluctuations causes significant cancellations in the 
denominator of Eq.~\ref{Eq:MonteCarlo} and renders the method ineffective due to large statistical uncertainties 
in $\langle \Omega \rangle$. 
In the literature, this problem is referred to as the Monte Carlo sign problem and it is common to Quantum Monte Carlo methods in fermionic many-body 
problems (see f.x. the review in Ref.~\cite{linden1992}).
 For any Hamiltonian (Eq.~\ref{Eq:GenHam}) with  all $V_\alpha<0$, $h_\sigma$ are always time-reversal invariant, since all $s_\alpha$ are real (Eq.~\ref{Eq:LinHam}). As was shown by Lang {\it et al.} \cite{lang1993}, time-reversal invariance of $h_\sigma$ implies that the eigenvalues of the matrix $\mathrm{U}_\sigma$ come in complex-conjugate pairs which, in turn, ensures that the grand-canonical partition function $\mathrm{Tr} U_\sigma$ is positive definite. In the canonical
ensemble, projections on even number of particles always preserve the good sign as long as the grand canonical partition function is positive definite. However for systems with odd-number of particles, projections onto an odd number of particles usually reintroduces the sign problem at large values of $\beta$
even when the grand canonical partition function is positive definite.

Although Quantum Monte Carlo simulations are susceptible to the sign problem for a general two-body interaction and require practical approaches
to avoid it~\cite{scalapino1981,chen2004,
carlson2003,giorgini2005,forbes2012,burovski2006,bulgac2006,magierski2009,endres2011,carlson2011,mukherjee2013},
  purely attractive two-body 
interactions are known to be free of this restriction~\cite{bulgac2006,chris2012}.  For the benefit of both nuclear and atomic physics communities, we 
also demonstrate the absence of the sign problem explicitly for an attractive zero-range interaction in the next section.

\section{Sign Properties of the Zero-Range Interaction}
We now consider the zero-range interaction in the $jj$-coupling scheme  which is discussed in full detail in appendix \ref{appa}. We write the two-body Hamiltonian in the so-called  {\it pairing} (or {\it particle-particle}) {\it decomposition}~\cite{lang1993,koonin1997} as
\begin{equation}
H_2=\frac{1}{2}\sum_{abcd}\sum_{JM} V_J(ab,cd) A^{\dagger}_{JM}(ab)A_{JM}(cd),
\end{equation}
where the pair operators are defined by
\begin{equation}
A_{JM}^{\dagger}(ab)=\sum_{m_a m_b}\langle j_a m_a j_b m_b | JM \rangle a^{\dagger}_{j_b m_b} a^{\dagger}_{j_a m_a}. 
\end{equation}

We now introduce the combined indices $i=(ab)$ and $j=(cd)$ to write $V_J(i,j)$ which is a symmetric matrix. Our goal is to diagonalize the matrix and inspect the signs of the eigenvalues. As demonstrated in \cite{lang1993}, the interaction will produce no sign problem when all of its eigenvalues are negative. Obviously the problem splits into blocks of given $J$, so we work in a fixed $J$ subspace. 

The crucial observation is that $V_J(ab,cd)$ can be factorized in the following way. Firstly, we define the following quantity:
\begin{widetext}
\begin{equation}
f_J(ab)\equiv\frac{1}{\sqrt{2}}(-1)^{l_a+j_b+1/2} [l_a][l_b][j_a][j_b] 
\left\{\begin{matrix} l_a & j_a & \tfrac{1}{2} \\ j_b &l_b&J\\ \end{matrix}\right\}
\left(\begin{matrix} l_a & l_b & J \\ 0&0&0\\ \end{matrix}\right)e^{i\theta}\sqrt{\frac{|V_0|}{4\pi}} r R_{n_a l_a}(r) R_{n_b l_b}(r),
\end{equation}
\end{widetext}
where $e^{2i\theta}=sgn(V_0)$ and $[j]=\sqrt{2j+1}$. Notice that $f_J(ab)e^{-i\theta}$ is a purely real number. In terms of the combined indices we now have 
\begin{equation}
V_J(i,j)=\int_{0}^{\infty} dr f_J(i)f_J(j).
\end{equation}
Since this matrix is real symmetric, there is a basis of orthonormal eigenvectors. Let us denote this basis ${u^k}$ and the corresponding eigenvalues ${\lambda^k}$. The dimension is given by the number of pairs in the given model space that can couple to total angular momentum $J$. Consider now for a given $k$ the product $(u^k)^T V_J u^k$, where $T$ denotes the transpose. Inserting the explicit form of $V_J$ we have
\begin{eqnarray}
&u^T V_J u=\sum_{ij} u^{k}(i) V_J(i,j) u^{k}(j)=&\nonumber\\&\int_{0}^{\infty} dr \left[  \sum_i f_J(i) u^{k}(i)\right]^2=\lambda^{k},&
\end{eqnarray}
where the last equality follows from the eigenvalue equation and the fact that $u^k$ is normalized. We thus see that the eigenvalues are equal to some real number squared times a phase $e^{2i\theta}=sgn(V_0)$. Therefore, the sign of $V_0$ is also the sign of the eigenvalues. We thus have the result that any attractive zero-range interaction ($V_0<0$) will have no sign problem, whereas the repulsive ($V_0>0$) case can never give a positive-definite path integral.

The simple form of $V_J(i,j)$ allows us to prove some further properties of its spectrum. Define (for fixed $J$ not shown) the row vector $f=[f_1 f_2\ldots f_n]$, where $n$ counts the pairs, such as to fulfill $V_J=\int dr f^T f$. Now pick a row vector orthogonal to $f$ so that $f g^T=0$. Then we see that $V_J g^T=\int dr f^T f g^T=\int dr f^T(fg^T)=0$, thus all vectors orthogonal to $f$ are in the null-space of $V_J$. We therefore have only one non-zero eigenvalue for each $J$ and $n-1$ eigenvectors with zero eigenvalue. The sole non-zero eigenvalue has the value $\int dr f f^T$ and the eigenvector $f^T$. We thus see that the zero-range pairing interaction has a very simple structure after diagonalization.

As mentioned, the above proof was carried out in the so-called pairing decomposition with the operators $A_{JM}^{\dagger}(ab)$ and $A_{JM}(cd)$. In many nuclear applications of the methods, the calculations are carried out in the density decomposition \cite{koonin1997}. However, the exact path integral is independent of the particular representation and the above result will still hold. In particular, the change from pairing to density decomposition is in practice a re-coupling of the angular momenta involved (and a change of the one-body terms that we are not concerned with). Since re-couplings corresponds to changes of basis the result for the eigenvalues still holds. In Appendix~\ref{appb}, we include a proof based on the 
$m$-scheme and the density decomposition for completeness.
We have also done explicit numerical checks of this fact and confirmed the general statement.

The good sign properties of the zero-range pairing rested on the fact that it could be factorized, which is more commonly referred to as separability of the zero-range interaction. A non-zero range interaction would not have this property and positive eigenvalues with associated sign problems can be expected. We note again that even with an interaction that has good sign properties, a system
with an odd number of particles will still have a sign problem at low temeperature \cite{koonin1997}.

\section{Pairing Correlations in SMMC}
To illustrate the above discussion, we now turn to an 
example of small Fermi systems and their pairing properties. 
The lack of sign problems for the zero-range interaction 
means that the SMMC can be applied. This was done
recently and the energetics and convergence properties have
been reported in Ref.~\cite{zinner2008}. Here we will focus on 
pairing properties which is another expectation value that 
is accessible through the SMMC method. The discussion above
in fact implies that the two-component Fermi system in a trap
can be mapped onto what is known a pure pairing problem due to 
the simple form of the interaction. Note that 
the strength of the two-body matrix is state-dependent. This is 
an important difference in comparison to typical models for 
large-scale two-component systems that are employed for instance in 
the BCS theory of conventional macroscopic superconductors.

In the basic Hamiltonian in Eq.~\eqref{ham}, we parametrize the 
interaction by $V_0$ which has units of energy times volume. As 
discussed in Ref.~\cite{zinner2008}, the interaction strength
can be written $V_0=-g\hbar\omega b^3$ (remember that we 
consider $V_0<0$ only to avoid sign issues), where $g$ is 
now a convenient dimensionless strength parameter. In experiments
on atomic gases the interactions are usually parametrized via the
two-body scattering length, $a$, which can be tuned by 
applied fields \cite{bloch2008}. Relating the value of $g$ to 
the value of $a$ is therefore crucial and will in general depend
on the model space used. Here we will consider $g$ to be a 
parametric quantity to describe pairing, but for the sake of completeness
we note that the values used below,  $g=10$ and $g=20$---in a 
model space  consisting of the major shells of $s+p+sd$--- 
correspond to $a/b=-1.0$ and $a/b=11$ respectively.

To develop a better understanding for the energetics of the pairing strength considered in this section,
we can consider a simple pairing model with the structure 
\begin{equation}
H=\sum_i G a_{i}^{\dagger}a_{i}+\frac{V}{2}\sum_{i,j}
a_{i}^{\dagger}a_{\bar{i}}^{\dagger}a_{\bar{j}}a_{j},
\end{equation}
where $i,j$ denotes single-particle levels and $\bar{i}$ is a 
time-reversed state. $G$ and $V$ are the level spacing and 
pairing strength respectively. In units of $G=1$,  
$V\sim 1$ is a regime  of competition between single-particle
excitations and pairing, while the regime of $V\sim 10$ is  
pairing dominated. The model we study here differs from the simple
pairing model by having state-dependent matrix elements given 
by the overlap of different oscillator single-particle states. 
However; we can still give an overall estimate of the typical 
matrix elements in units of the single-particle 
level spacing, $\hbar\omega$. In the regime characterized by $g=10$, magnitude of 
a typical matrix element is of the order 1 and in the light of the simple model mentioned above, 
we expect  pairing and level structure to be in competition. In comparison, the regime described by $g=20$ should 
naturally be pairing dominated. These ascertainments are perfectly
consistent with the typical discussion of BCS-BEC crossover 
\cite{bloch2008,giorgini2008} when considering the 
corresponding values of $a/b$ cited above. Thus, we expect $a/b<0$ 
and $a/b>0$ to be in the weak and strong pairing regimes, respectively.

\begin{figure*}[ht!]
\begin{center}
\includegraphics[scale=0.50,clip=true]{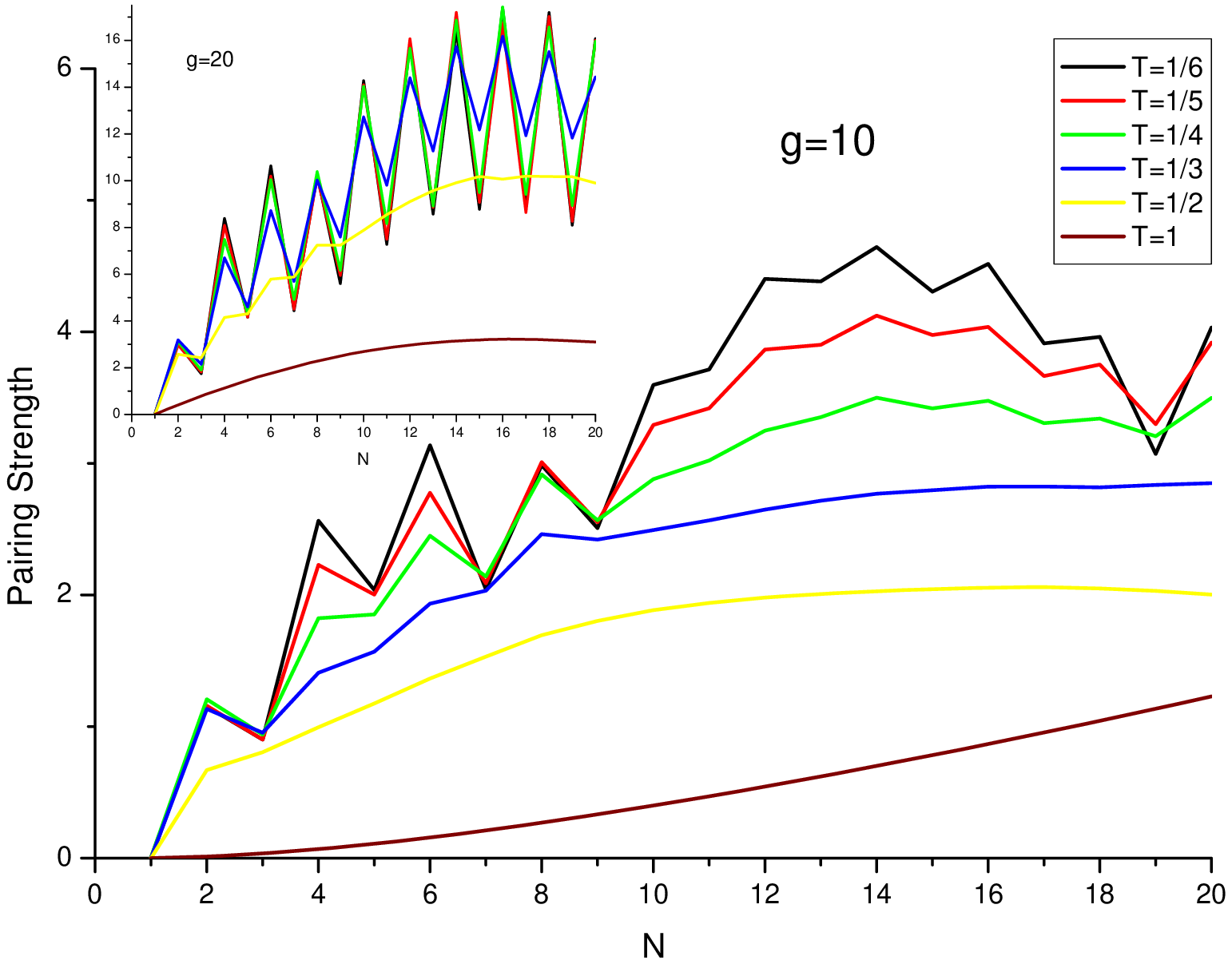}
\caption{(color online) The pairing strength as a function of particle number, $N$, for various
temperatures, $T$ (in units of $\hbar\omega$) for $g=10$. The upper left inset shows the
results for $g=20$. The uncertainties are very small and not shown.}
\label{fig2}
\end{center}
\end{figure*}

To study the pairing properties, we consider the expectation value of a
number-conserving BCS-like pair matrix
\begin{equation}
M_{\alpha,\alpha'}=\langle \Delta^\dagger(j_a,j_b)\Delta(j_c,j_d)\rangle,
\label{pairmat}
\end{equation}
with the $J=0$ pair operator
\begin{equation}
\Delta^\dagger=\frac{1}{ \sqrt{1+\delta_{ab}} }
\left[ a_{j_a}^{\dagger} \times a_{j_b}^{\dagger} \right]^{JM=00}
\end{equation}
where $a_{j_a}^{\dagger}$ creates a particle in orbit $j_a$ (which is the combination of orbital and spin
angular momentum of the fermions). This operator is thus a measure of the pairing content corresponding to $J=0$.
An indication of the pairing correlations can be
obtained from the sum over all matrix elements, defining the
pairing strength in the following \cite{ozen2007}.
Since we employ a finite temperature 
formulation of the SMMC method, we, however, 
need to eliminate the thermal correlations that would 
be present in the non-interacting system. We therefore 
subtract the 'mean-field' values---calculated at the same $T$ but with $g=0$---to obtain the genuine pairing correlations.

The pairing correlations are important in nuclear
physics in several respects. A particular example is the influence
of pairing on nuclear level density distributions~\cite{ozen2007}
which are crucial for addressing nuclear reactions of astrophysical 
interest \cite{friedel1997}. In cold atomic gases, pairing correlations
are observable in what is usually called noise correlations
\cite{altman2004}. These two-point correlations have been measured 
in experiments using optical lattice potentials and are employed to demonstrate  
bunching for bosonic \cite{folling2005} and anti-bunching for fermionic
atoms \cite{rom2006}. The pairing correlations we consider here should
therefore be directly measurable in the cold atomic gases. Alternatively,
a projection method can be used, wherein one rapidly changes the interaction
strength to convert all pairs into molecules \cite{rapidramp}. The 
momentum distribution of the molecules can subsequently be measured by
turning the trapping potential off, and this carries the imprint of the 
original many-body state in the trap prior to molecular conversion and 
release of the system.

In Fig.~\ref{fig2} we show the pairing strength as a function
of particle number for different temperatures.
The most striking feature is naturally the odd-even staggering. The
relative reduction of pairing strength for odd-particle numbers is related
to the blocking of scattering of pairs into the orbital occupied by the
unpaired particle. The non-interacting systems have closed-shell configurations
for $N=2,8,20$. With interaction switched on, these configurations manifest
themselves by a relative reduction of the pairing strength (overlaid
by a general increase due to a growing number of pairs) and a larger
resistance against temperature increase. The strong dips observed for
particle numbers $N=7$, 9, and 19 are also connected to the shell closures.
Relatedly, the pairing strength is largest for mid-shell systems. In the inset
one can see that the staggering is larger for $g=20$ and persists to larger
temperatures as expected.

\begin{figure*}[ht!]
\begin{center}
\includegraphics[scale=0.50,clip=true]{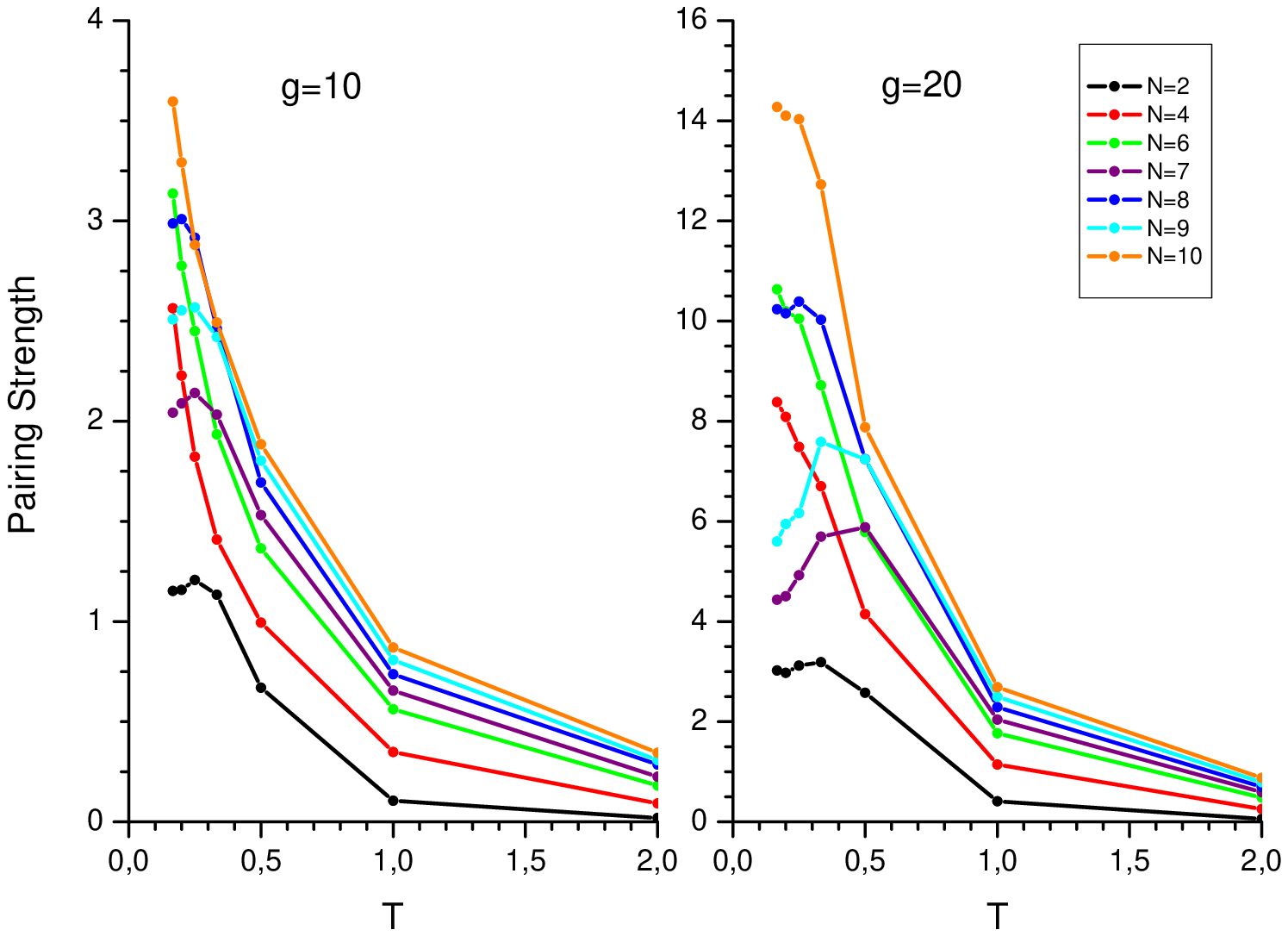}
\caption{(color online) The pairing strength 
as a function of temperature, $T$ (in units of $\hbar\omega$), for various
particle numbers, $N$ and for $g=10$ (left) and $g=20$ (right). 
Uncertainties are small and not shown. Note the different  scales
in the two panels.}
\label{fig3}
\end{center}
\end{figure*}

To investigate further the transition between a paired state
and a normal state, we show the pairing
strength for $g=10$ and $g=20$ as a function of $T$ for 
selected particle numbers in Fig.~\ref{fig3}.
We note that, in the high temperature regime, pairing 
correlations are ordered with increasing number
of particles (an indication of the equipartitioning in the model space) 
and that they go through a rapid
and monotonous decay. In contrast, the low temperature regime is 
dominated by structure and odd-even effects. 
Notice the persistence in the pairing strength in the systems 
with $N=2$ and 
$N=8$ (also the case for $N=20$, which is not shown) due to 
the shell closures. Shown are also the cases of $N=7$ and $N=9$,
which exhibit large dips in the pairing strength
in Fig. \ref{fig2}. They are generally below the neighboring 
even-$N$ systems at low $T$, yet again
confirming that an unpaired particle has a significant
blocking effect on the pairing strength. Furthermore; they have 
the same structure as the neighboring closed
shell $N=8$, but at lower
magnitudes. It is also interesting to observe that, for systems with $N=7$ and $9$, the
pairing strength is largest at finite $T$, reflecting the competition
between blocking by the unpaired particle and thermal excitations
which moves the unpaired particle across the shell closure reducing
the blocking effect.
A similar effect has been found in the SMMC
studies for nuclei with odd-nucleon numbers~\cite{Engel98}.
Comparing the $g=10$ and $g=20$ results, we see that the above effects are
more pronounced for the stronger pairing strength and persists to higher temperature.
This is consistent with the discussions above. Similar evidence for a transition at finite $T$ in
a homogeneous system in both energy and pair correlation was found in \cite{akkineni2007}.

\subsection{Connection to other pairing phenomena}
Many pairing studies consider only pairs of particles in time-reversed 
states with an attractive zero-range interaction of constant magnitude $g<0$. This 
is, for instance, the case in condensed-matter physics when appyling the simplest
version of the BCS pairing theory to a homogeneous Fermi gas with an attractive interaction 
in relative momentum zero and spin singlet states 
($\bm k,\uparrow$ pairs with $-\bm k,\downarrow$ only). This philosophy of pairing
time-reversed states can be continued to non-homogeneous systems but at the price
of getting a state-dependent gap function, $\Delta_i$, in general, where $i$ denotes
the mean-field single-particle levels that are subjected to a pairing interaction.
(the mean-field could arise from a Hartree-Fock calculation).

In nuclear physics, pairing models often employ this restriction, as in the case of 
the pairing force problem (see \cite{fetter1971}) 
which has the property that it is exactly solvable. A justification for
these models comes from the fact that the pairing force usually has a short-range
and for two nucleons in a single mean-field level, the total $J=0$ pairs have the 
strongest gain in binding \cite{fetter1971}. In this single level
case, these pairs are built from time-reversed states \cite{brink1993}.

If we consider the case of cold atomic gases, we start from the zero-range interaction
and an external trap providing the mean-field. In a BCS picture, this implies
that we have general matrix elements (as given in Eq.~\eqref{LSJres}) and a 
state-dependent gap, $\Delta_i$. However, there are now different regimes of 
interest depending on the strength, $g$, and the level spacing, $\hbar\omega$. This 
has been discussed in Ref.~\cite{bruun2002} using the Bogoliubov-de Gennes equations (more commonly
called the Hartree-Fock-Bogoliubov equations in  nuclear physics) along with the local density 
approximation to describe larger systems. There it was found that an intra- and an 
inter-shell pairing regime appears, depending on whether the typical gap parameter
satisfies $\Delta<\hbar\omega$ (intra) or $\Delta>\hbar\omega$ (inter). Since the zero-range
interactions which are employed in the Bogoliubov-de Gennes approach are precisely 
time-reversed, one has $l_a=l_b$ and $l_c=l_d$ in Eq.~\eqref{LSJres}.

\begin{figure}[ht!]
\begin{center}
\includegraphics[scale=0.43,clip=true]{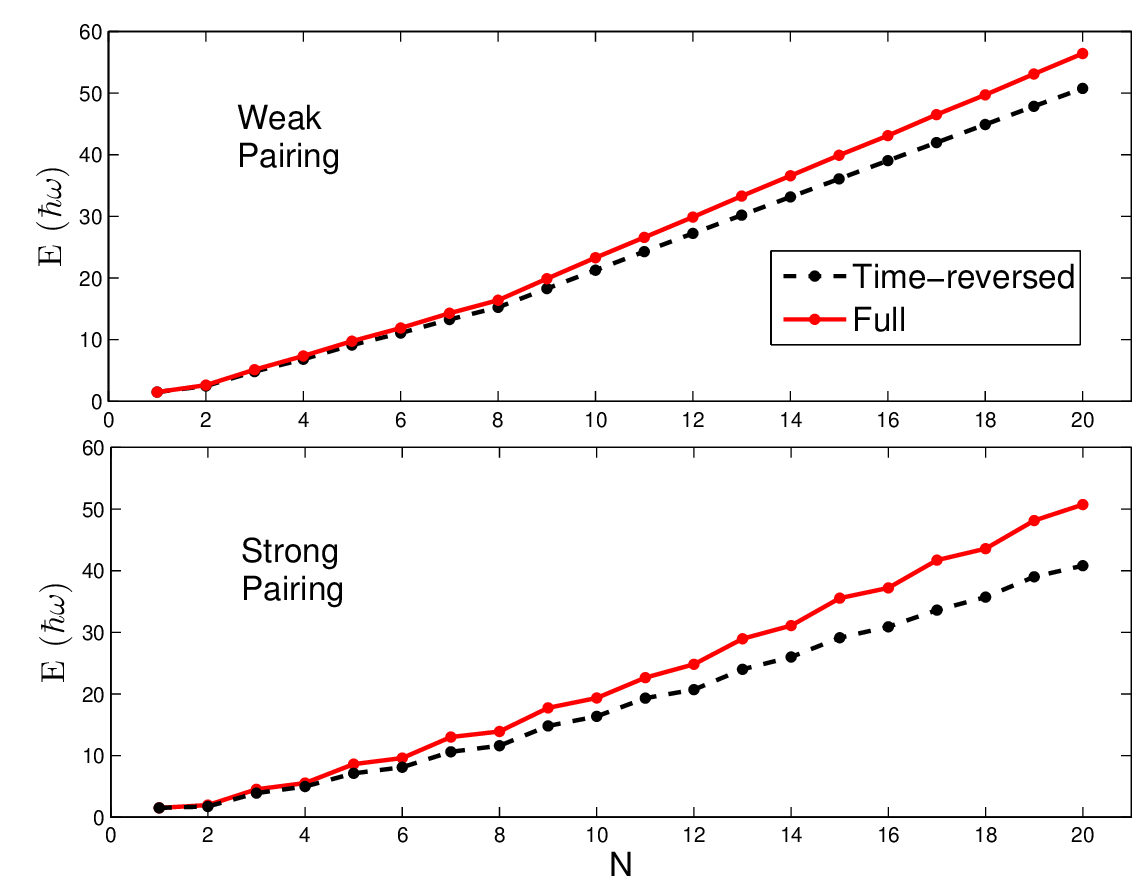}
\caption{(color online) The energy in units of $\hbar\omega$ of 
as function of particle number $N$ for time-reversed only (dashed line) and full (solid line) 
interaction as discussed in the text. Upper panel is for strength $g=10$ (weak pairing), while the lower one
has $g=20$ (strong pairing).}
\label{fig4}
\end{center}
\end{figure}

Here we are concerned with small systems, and it is clear that the mean-field Bogoliubov-
de Gennes should break down as particle numbers become small, and correlations beyond the
mean-field are strong. In order to get a quantitative feeling for these additional correlations
we can compare a model where only time-reversed states are used in the 
interaction ($l_a=l_b$ and $l_c=l_d$) and the full zero-range interaction where
all states that give non-zero contributions to Eq.~\eqref{LSJres} are taken into 
account. It can be readily observed that our proof of good sign properties will
hold in both cases (time-reversed states are a special case)
and the SMMC should work perfectly well. 

In Fig.~\ref{fig4}, we plot the energy of systems with particle numbers of $N=1-20$ for 
two kinds of interaction; one that pairs only the time-reversed states (dashed line) and 
the full zero-range interaction (solid line) in both
the weak ($g=10$ in the upper panel) and the strong pairing ($g=20$ in the lower panel) regimes.
In general, we see that the full interaction gives a somewhat higher energy
than the time-reversed one. This is most likely caused by the fact that the 
full interaction allows low-lying pairs to correlate with pairs in higher
shells and thus raise the energy. We see that both interactions capture
the shell effect at $N=8$, while the full interaction seems to produce more 
structured odd-even effects due to strong pairing. In the overall, however, we do not
observe a pronounced difference between a pairing interaction involving only 
time-reversed states and the full zero-range pairing interaction,
the latter being the physical interaction employed in studies of ultracold atomic Fermi gases.
Our findings thus indicate that pairing involving only time-reserved states can 
be a good approximation for the study of small systems as well. Of course, we have to 
stress that in this limit the shell structure effects are very important and we do
not expect this to be captured accurately by local-density approximations; thus, 
the full discrete external trap spectrum must be considered.

\section{Summary and outlook}
Studies of small two-component Fermi systems in tight external traps are 
currently being pursued experimentally \cite{selim1,selim2} in the realm
of cold atomic gas physics. 
Here we demonstrate how the mapping of
the atomic system to an equivalent problem in nuclear physics can be achieved.
It has the important feature that there is no sign problem associated with the typical 
choice of a zero-range interaction within the grand-canonical formulation of the SMMC approach. 
As we have discussed, the atomic 
interaction between the two internal hyperfine states is more general than 
the typical pairing force used in many investigations, and it was 
therefore not {\it a priori} clear that the corresponding nuclear SMMC
problem would be free of the sign problem. 
The alternative approach of using large-scale shell-model diagonalization
however, is computationally challenged by the number of configurations which grows exponentially with the model 
space size; in contrast,  the size of the problem scales only quadratically in the SMMC approach~\cite{koonin1997}. 
Truncation of the model space may be used to reduce the size of the problem to a certain extent. 
In low-dimensional systems, which are 
currently under intense study in atomic physics, the reduced size of the matrix problem may allow
a direct diagonalization of the many-body Hamiltonian (a recent pairing study using 
nuclear-inspired methods can be found in Ref.~\cite{rontani2009}). However; in the full three-dimensional case,
the SMMC method seems to be the only tractable approach at the moment. 

We note that there is growing 
interest in multi-component Fermi systems in atomic gas physics. Three-component 
mixtures of $^{6}$Li have been realized a few years ago and continue to 
be a hot topic \cite{nakajima2011}. Fermionic systems with four or more components 
are also being pursued since it is possible to realize such systems by using
not alkali but rather alkali-earth atoms which can have many degenerate hyperfine
states, allowing the realization of many interesting models of magnetism and pairing \cite{earth}. 
From a nuclear physics point-of-view, a multi-component system can be mapped onto the 
isospin degree of freedom. In the case of four-component Fermi systems, one should 
therefore be able to perfectly map the problem onto the isospin $1/2$ times spin $1/2$
formalism and exploit the corresponding advanced calculational tools available 
in nuclear physics.

In closing, we would also like to point out that spin-orbit coupling has recently
become a heavily pursued topic in ultracold atomic systems since it is now
possible to implement by optical means for both bosonic \cite{lin2009,lin2011} and fermionic 
atomic systems \cite{wang2012,cheuk2012}. These studies produce a spin-orbit coupling 
of the kind used in mostly condensed matter and solid state, which has the form
of ${\bm s}\cdot {\bm k}$, i.e. of a spin-linear momentum coupling. However, it was
recently shown that it is possible to use applied optical fields that impart 
orbital angular momentum instead of linear momentum on atoms \cite{moulder2012,beattie2012}.
It should thus be within reach to create terms that are similar to the traditional 
spin-orbit term encountered in nuclear physics, i.e. of the form of $\bm s\cdot \bm l$.
This would  immediately imply that the $jj$-coupling be the more suitable 
approach for the study of small atomic Fermi systems with optically induced 
spin-orbit interactions. Since the external laser intensity is typically 
a multiplicative factor on the coupling terms, we expect that one can correspondingly
address the full range of spin-orbit strength from weak to strong, both 
experimentally and theoretically.

\emph{ Note added after completion:} A related study of small Fermi systems using 
a method very similar to the one discussed here has been presented in Ref.~\cite{chris2012}.
That study considers pairing correlations defined in a similar fashion to our Eq.~\ref{pairmat}.

\section{Acknowledgments}
We acknowledge fruitful discussions with Karlheinz Langanke, David Dean, Klaus M\o lmer and Christopher Gilbreth. C\"O thanks Thomas Pappenbrock for suggestions on the current work. NTZ would like to thank Niels Leth Gammelgaard, Thomas Kragh, and Mark S. Rudner for enlightening discussion on some linear algebraic details, and David Pekker for reading and commenting on an early draft. We thank the referees for comments and 
suggestions that have improved the presentation and discussion.

\begin{appendix}
\section{The Zero-range Force in the $jj$-coupling Scheme}\label{appa}
To make explicit the rotational invariance in nuclear applications, matrix elements of the two-body interaction are often specified in the $jj$-coupling scheme by
\begin{widetext}
\begin{equation}
V_J(ab,cd)=\langle [\psi_{j_a}(\vec{r}_1)\times\psi_{j_b}(\vec{r}_2)]^{JM} |V(\vec{r}_1,\vec{r}_2) |  
[\psi_{j_c}(\vec{r}_1)\times\psi_{j_d}(\vec{r}_2)]^{JM} \rangle,
\label{vj}
\end{equation}
\end{widetext}
where $a,b,c$ and $d$ denote single-particle orbitals and $j_a,j_b,j_c,$ and $j_d$ are their respective angular momenta. Notice that $V_J$ is independent of the total projection $M$ (as can be seen by applying the Wigner-Eckart theorem). In analogy with the nuclear shell model, single-particle orbitals associated with an external mean field (here assumued to be spherical) carry the quantum numbers $(nlm_l)$ and the internal (spin-half) quantum numbers $(\tfrac{1}{2}m_s)$. The external and internal angular momenta can be coupled through $\vec{j}=\vec{l}+\vec{s}$ to give the total angular momentum $j=l\pm 1/2$ for a given single-particle orbital.

As discussed in the main text, the zero-range interaction we employ connects only two-body states with spin-singlet internal states; $|S=0,M_s=0\rangle$. To this end, it is more convenient to transform the  $jj$-coupling scheme to the $LS$-coupling scheme. This can easily be achieved using the standard techniques of angular momentum \cite{brink1993}:
\begin{widetext}
\begin{eqnarray}
|(l_a s_a)j_a, (l_b s_b)j_b,JM \rangle = \sum_{L,S} [L][S][j_a][j_b]\left\{\begin{matrix} l_a & s_a & j_a \\ l_b&s_b&j_b\\L&S&J
\end{matrix}\right\}
|(l_a l_b)L,(s_a s_b)S,JM \rangle.
\label{9jinv}
\end{eqnarray}
\end{widetext}
Here we are interested in the $s_a=s_b=1/2$ case, and, since the interaction contains a projection onto spin singlet states, only need the $S=0$ component of this transformation. Using a reduction on the $9j$ symbol \cite{brink1993}, the projection can be written
\begin{widetext}
\begin{eqnarray}
&P_{S=0}|(l_a \tfrac{1}{2})j_a, (l_b \tfrac{1}{2})j_b,JM \rangle = \sum_{L} [L][j_a][j_b]
\left\{\begin{matrix} l_a & \tfrac{1}{2} & j_a \\ l_b&\tfrac{1}{2}&j_b\\L&0&J
\end{matrix}\right\}
|(l_a l_b)L,(\tfrac{1}{2} \tfrac{1}{2})0,JM \rangle=\nonumber\\&
(-1)^{L+l_a+j_b+2j_a-1/2}\frac{[j_a][j_b]}{\sqrt{2}}
\left\{\begin{matrix} L & j_a & j_b \\ 1/2&l_b&l_a\\ \end{matrix}\right\}\delta_{LJ}
|(l_a l_b)L,(\tfrac{1}{2} \tfrac{1}{2})0,JM \rangle,&
\end{eqnarray}
\end{widetext}
where $P_{S=0}=(1-{\bm \sigma}_1\cdot{\bm \sigma}_2)/4$ is the projection onto the spin singlet state.
The remaining zero-range interaction of course only acts on the external quantum states, thus we have to evaluate matrix elements between coupled states with operators acting on only one of the degree of freedom. Since the spin part is trivial for singlets we simply have the result (keeping both $L$ and $J$ for clarity even though $L=J$)
\begin{eqnarray}
&\langle(l_a l_b)L,(\tfrac{1}{2} \tfrac{1}{2})0,JM | V(\vec{r}_1-\vec{r}_2) |(l_c l_d)L,(\tfrac{1}{2} \tfrac{1}{2})0,JM \rangle=&\nonumber\\
&\langle(l_a l_b) LM | V(\vec{r}_1-\vec{r}_2)|(l_c l_d) LM\rangle,&
\end{eqnarray}
where we have explicitly indicated the orbital angular momenta of all states involved. For the zero-range interaction 
$V(\vec{r}_1-\vec{r}_2)=V_0\delta(\vec{r}_1-\vec{r}_2)$, the latter matrix element can be found in many textbooks (see for instance \cite{brink1993}) and
is given by
\begin{widetext}
\begin{eqnarray}
&\langle l_1l_2JM|V_0\delta(\vec{r}_1-\vec{r}_2)|l'_1l'_2J'M'\rangle =&\nonumber\\&\delta_{J,J'}\delta_{M,M'} [l_1][l_2][l'_1][l'_2] 
\left(\begin{matrix}l_1&l_2&J\\0&0&0 \end{matrix}\right)
\left(\begin{matrix}l'_1&l_2'&J\\0&0&0 \end{matrix}\right)
\frac{V_0}{4\pi}
\int_{0}^{\infty}dr r^2 R_{n_1l_1}(r) R_{n_2l_2}(r) R_{n'_1l'_1}(r) R_{n'_2l'_2}(r).&
\label{general}
\end{eqnarray}
\end{widetext}

We can now insert all these formulae into eq.~\ref{vj} to get an expression for the $J$-scheme interaction:
\begin{widetext}
\begin{eqnarray}
&V_J(ab,cd)=\delta_{J,L}(-1)^{l_a+l_c+2j_a+2j_c+j_b+j_d-1}\frac{[j_a][j_b][j_c][j_d]}{2}
\left\{\begin{matrix} L & j_a & j_b \\ \tfrac{1}{2} &l_b&l_a\\ \end{matrix}\right\}&\nonumber\\
&\left\{\begin{matrix} L & j_c & j_d \\ \tfrac{1}{2} &l_d&l_c\\ \end{matrix}\right\}\langle(l_a l_b) L0 | V(\vec{r}_1-\vec{r}_2)|(l_c l_d) L0\rangle=
(-1)^{j_b+j_d+l_a+l_c+1}\frac{[j_a][j_b][j_c][j_d]}{2}
\left\{\begin{matrix} J & j_a & j_b \\ \tfrac{1}{2} &l_b&l_a\\ \end{matrix}\right\}
\left\{\begin{matrix} J & j_c & j_d \\ \tfrac{1}{2} &l_d&l_c\\ \end{matrix}\right\}&\nonumber\\
&[l_a][l_b][l_c][l_d] 
\left(\begin{matrix}l_a&l_b&J\\0&0&0 \end{matrix}\right)
\left(\begin{matrix}l_c&l_d&J\\0&0&0 \end{matrix}\right)
\frac{V_0}{4\pi}
\int_{0}^{\infty}dr r^2 R_{n_al_a}(r) R_{n_bl_b}(r) R_{n_cl_c}(r) R_{n_dl_d}(r),&
\label{LSJres}
\end{eqnarray}
\end{widetext}
where the second equality comes from using the formula in eq.~\ref{general}. Notice that the phase can be written with $l_b+l_d$ instead of $l_a+l_c$ since $l_a+l_b+l_c+l_d$ is even due to the restrictions from the Clebsch-Gordon coefficients.
This is the general interaction in the spin singlet state and $l_a$, $l_b$, $l_c$, and $l_d$ can in general be different as long as they couple pairwise to $L=J$. For the pairing interaction in the time-reversed states discussed in the text, we have $l_a=l_b$ and $l_c=l_d$.

The formula above explicitly shows that $l_a+l_b+J$ and $l_c+l_d+J$ must be even. However, the multipole expansion used to arrive at this expression implicitly requires that also $l_a+l_c+J$ and $l_b+l_d+J$ be even. Notice also that the factor $l_a+l_c$ means that pairing across two opposite parity major shells can be repulsive for $V_0<0$. This is well-known in nuclear pairing studies \cite{koonin1997}.

As discussed in \cite{koonin1997}, the physical matrix elements used in the nuclear shell model must be antisymmetrized. This can be achieved by using the definition
\begin{eqnarray}
&V_{J}^{A}(ab,cd)=\frac{1}{\sqrt{(1+\delta_{ab})(1+\delta_{cd})}}&\nonumber\\&\left[V_J(ab,cd)-(-1)^{j_c+j_d-J}V_J(ab,dc)\right].&
\end{eqnarray}
However, as by angular momentum algebra one may show that $V_J(ab,dc)=-(-1)^{j_c+j_d-J}V_J(ab,cd)$. Thus we have the simple result
\begin{equation}
V_{J}^{A}(ab,cd)=\frac{2}{\sqrt{(1+\delta_{ab})(1+\delta_{cd})}}V_J(ab,cd).
\label{anti}
\end{equation}
This is not surprising since we argued that only the antisymmetric $S=0$ spin-singlet component should have non-zero matrix elements. We have effectively enforced the Pauli principle in this manner.

\section{Sign properties in the $m$-scheme}\label{appb}
It is also possible to demonstrate that the contact interaction is free of the sign problem in the 
so-called {\it density decomposition}~\cite{koonin1997}. For convenience, we define the single-particle states
by $\vert nlms\sigma\rangle\equiv\vert am\sigma\rangle$ where
$a=(nl)$ and $\sigma=\pm 1/2$.  In the $m$-scheme approach, 
the two-body part of the Hamiltonian (we omit the one-body term without loss of generality) can be written as
\begin{eqnarray}
  H_2=\frac{1}{2}\sum_{{ \begin{array}{c}
 abcd \\ mm'\sigma\sigma'
\end{array}}} V_{abcd} a^\dagger_{am\sigma} 
    a^\dagger_{bm'\sigma'} a_{dm'\sigma'} a_{cm\sigma}
\end{eqnarray}
where we used the spin-independence of the interaction explicitly. Also notice that the contact interaction should not change the $m$-quantum number of the orbital angular momentum states, hence $V_\alpha$ are labelled by only the $nl$-quantum numbers of the states.
$H_2$ can be brought into the  density decomposition by a rearrangement of the creation and annihilation operators. The new two-body Hamiltonian (up to an additional one-body term), is now written as
\begin{widetext}
\begin{eqnarray}
  H'_2=\frac{1}{2}\sum_{{ \begin{array}{c}
 abcd \\ mm'\sigma\sigma'
\end{array}}} V_{abcd} a^\dagger_{am\sigma} 
    a_{cm\sigma} a_{bm'\sigma'}^\dagger a_{dm'\sigma'}=\frac{1}{2}\sum_{ij}V_{ij} \rho_i \rho_j
\end{eqnarray}
\end{widetext}
where $i=(ac)$ indicate two-body (particle-hole) indices and $\rho_i=\sum_{m\sigma} a^\dagger_{am\sigma}a_{cm\sigma}$ are density operators. Since $V_{ij}$ is a real symmetric matrix, it can 
be diagonalized by an orthogonal transformation
\begin{eqnarray}
   V_{ij}=\sum_\alpha \lambda_\alpha O_{\alpha i} O_{\alpha j}
\end{eqnarray}
Using this expression, $H'_2$ can be brought into a quadratic form
\begin{eqnarray}
  H'_2=\frac{1}{2}\sum \lambda_\alpha P_\alpha^2
\end{eqnarray}
where $P_\alpha=\sum_\alpha O_{\alpha i} \rho_i$. To manifest the time-reversal properties,  we now express $H'_2$ in a similar form to the two-body term in
Eq.~\ref{Eq:GenHam}. To this end, we introduce the 
time-reversed creation and annihilation operators in the Condon-Shortley phase convention (where the state $\vert lm \rangle$
is defined in terms of the spherical harmonics without the $i^l$ factor):
\begin{widetext}
\begin{eqnarray}
  {\bar{a}}^\dagger_{nlm\sigma} & = & (-1)^l (-1)^{l+m+1/2+\sigma} a^\dagger_{nl-m-\sigma}=(-1)^{m+1/2+\sigma} a^\dagger_{nl-m-\sigma} \\
  {\bar{a}}_{nlm\sigma} & = &  (-1)^l (-1)^{l+m+1/2+\sigma} a_{nl-m-\sigma}= (-1)^{m+1/2+\sigma} a_{nl-m-\sigma}
\end{eqnarray}
Using these, we obtain 
\begin{equation}
   {\bar{\rho}}_i=\sum_{m\sigma} {\bar{a}}^\dagger_{am\sigma} {\bar{a}}_{cm\sigma}= \sum_{m\sigma} (-1)^{2(m+1/2+\sigma)} a^\dagger_{a-m-\sigma} a_{c-m-\sigma}=\rho_i,
\end{equation}
\end{widetext}
from which it follows that ${\bar{P}}_\alpha=P_\alpha$. Hence we can write
\begin{equation}
   H'_2=\frac{1}{4}\sum_\alpha \lambda_\alpha \left\lbrace P_\alpha, {\bar{P}}_\alpha\right\rbrace.
\end{equation}
The condition for a good-sign interaction, that $h_\sigma$ (Eq.~\ref{Eq:LinHam}) is time-reversally invariant, demands that all $\lambda_\alpha<0$. For the contact interaction,  
$v=-g\delta(\mathbf{r}-\mathbf{r'})$, the matrix $V_{ij}$ is  negative-definite since 
\begin{equation}
V_{ij}=V_{abcd}\sim -g \int dr r^2 R_{nl}^2(r) R_{n'l'}^2(r)<0,
\end{equation}
thus the good-sign property is established.

\end{appendix}

\end{document}